\documentclass[twocolumn,conference]{IEEEtran}
\usepackage[T1]{fontenc}
\usepackage[latin9]{inputenc}
\usepackage{geometry}
\usepackage{dsfont}
\usepackage{amsmath}
\usepackage{amssymb}
\usepackage{graphicx}
\usepackage{setspace}
\usepackage[unicode=true,
 bookmarks=true,bookmarksnumbered=true,bookmarksopen=true,bookmarksopenlevel=1,
 breaklinks=false,pdfborder={0 0 0},pdfborderstyle={},backref=false,colorlinks=false]
 {hyperref}
\hypersetup{pdftitle={Your Title},
 pdfauthor={Your Name},
 pdfpagelayout=OneColumn, pdfnewwindow=true, pdfstartview=XYZ, plainpages=false}

\makeatletter
\usepackage[caption=false,font=footnotesize]{subfig}
\usepackage{cite}

\makeatother

\begin{document}
\title{Joint Scattering Environment Sensing \\and Channel Estimation for
\\Integrated Sensing and Communication}
\author{Wenkang Xu, Yongbo Xiao, An Liu and Minjian Zhao \\College of Information
Science and Electronic Engineering, Zhejiang University\\}
\maketitle
\begin{abstract}
This paper considers an integrated sensing and communication system,
where some radar targets also serve as communication scatterers. A
location domain channel modeling method is proposed based on the position
of targets and scatterers in the scattering environment, and the resulting
radar and communication channels exhibit a partially common sparsity.
By exploiting this, we propose a joint scattering environment sensing
and channel estimation scheme to enhance the target/scatterer localization
and channel estimation performance simultaneously. Specifically, the
base station (BS) first transmits downlink pilots to sense the targets
in the scattering environment. Then the user transmits uplink pilots
to estimate the communication channel. Finally, joint scattering environment
sensing and channel estimation are performed at the BS based on the
reflected downlink pilot signal and received uplink pilot signal.
A message passing based algorithm is designed by combining the turbo
approach and the expectation maximization method. The advantages of
our proposed scheme are verified in the simulations.
\end{abstract}

\begin{IEEEkeywords}
Integrated sensing and communication, location domain, scattering
environment sensing, channel estimation.
\end{IEEEkeywords}

\section{Introduction}

Radar sensing and wireless communication systems have been developed
independently for decades, and they are usually designed separately.
However, there are many similarities between sensing and communication
systems, such as signal processing algorithms, hardware architecture
and channel characteristics \cite{LiuFan_survey_signal_process,LiuFan_survey_dual_function}.
The sensing and communication functionalities are expected to mutually
assist each other by leveraging these similarities. 

We focus on the scattering environment in massive multi-input multi-output
(MIMO) Orthogonal Frequency Division Multiplexing (OFDM) integrated
sensing and communication (ISAC) systems, which reflects an interesting
similarity between radar sensing and communication in terms of channel
characteristics. The scattering environment includes two subsets,
i.e., radar targets and communication scatterers. However, some radar
targets also serve as communication scatterers in many cases. In an
ISAC scenario for vehicle networks, for instance, the BS needs to
localize vehicles and obstacles on the road and broadcast the sensing
data to every vehicle to realize automatic obstacle avoidance and
route planning. In this case, some vehicles and obstacles also contribute
to communication paths for neighboring vehicles. Due to the partial
overlap between radar targets and communication scatterers, radar
and communication channels will exhibit a partially common sparsity
in some sparse domains.

Recently, some related works have also exploited this similarity to
achieve target sensing and channel estimation in ISAC systems. In
\cite{LiuFan_JRC1}, based on the assumption that targets also serve
as scatterers for the communication signal, the authors proposed a
novel target sensing and channel estimation scheme. However, the target
sensing and channel estimation were carried out independently. In
\cite{Huangzhe_JRC2}, the authors merged target sensing and channel
estimation into a single procedure under the assumption that radar
targets and communication scatterers partially overlapped. The authors
in \cite{Wan_TSCE_UAV} studied an application of ISAC for unmanned
aerial vehicle (UAV) networks, in which a UAV communicated with the
terrestrial station while other UAVs and obstacles were viewed as
radar targets. In \cite{Gaudio_TSCE_OTFS1,Gaudio_TSCE_OTFS2}, each
radar target was also a communication receiver, and a two-step approach
was proposed to estimate the target location and the line-of-sight
(LoS) channel path. However, to the best of our knowledge, the existing
works did not consider joint scattering environment sensing and channel
estimation in MIMO-OFDM ISAC systems, where scattering environment
sensing refers to the joint localization of radar targets and communication
scatterers.

In this paper, we consider a time-division duplex (TDD) massive MIMO-OFDM
ISAC system and propose a joint scattering environment sensing and
channel estimation scheme. A novel location domain sparse representation
of radar and communication channels is introduced, which is suitable
to perform the joint localization of radar targets and communication
scatterers. The partially common sparsity of the location domain channels
is exploited to improve both target/scatterer localization and channel
estimation performance. However, some non-ideal factors, such as time
offset and user localization error, seriously degrade sensing and
estimation performance. In order to mitigate their impact, the accurate
estimation of these non-ideal factors is taken into account in the
algorithm design. A scattering environment aware turbo sparse Bayesian
inference (SEA-Turbo-SBI) algorithm is designed to solve the problem
by combining the turbo approach and the expectation maximization (EM)
method. 

\section{System Model}

\subsection{Architecture of the ISAC System}

\begin{figure}[h]
\begin{centering}
\includegraphics[width=90mm]{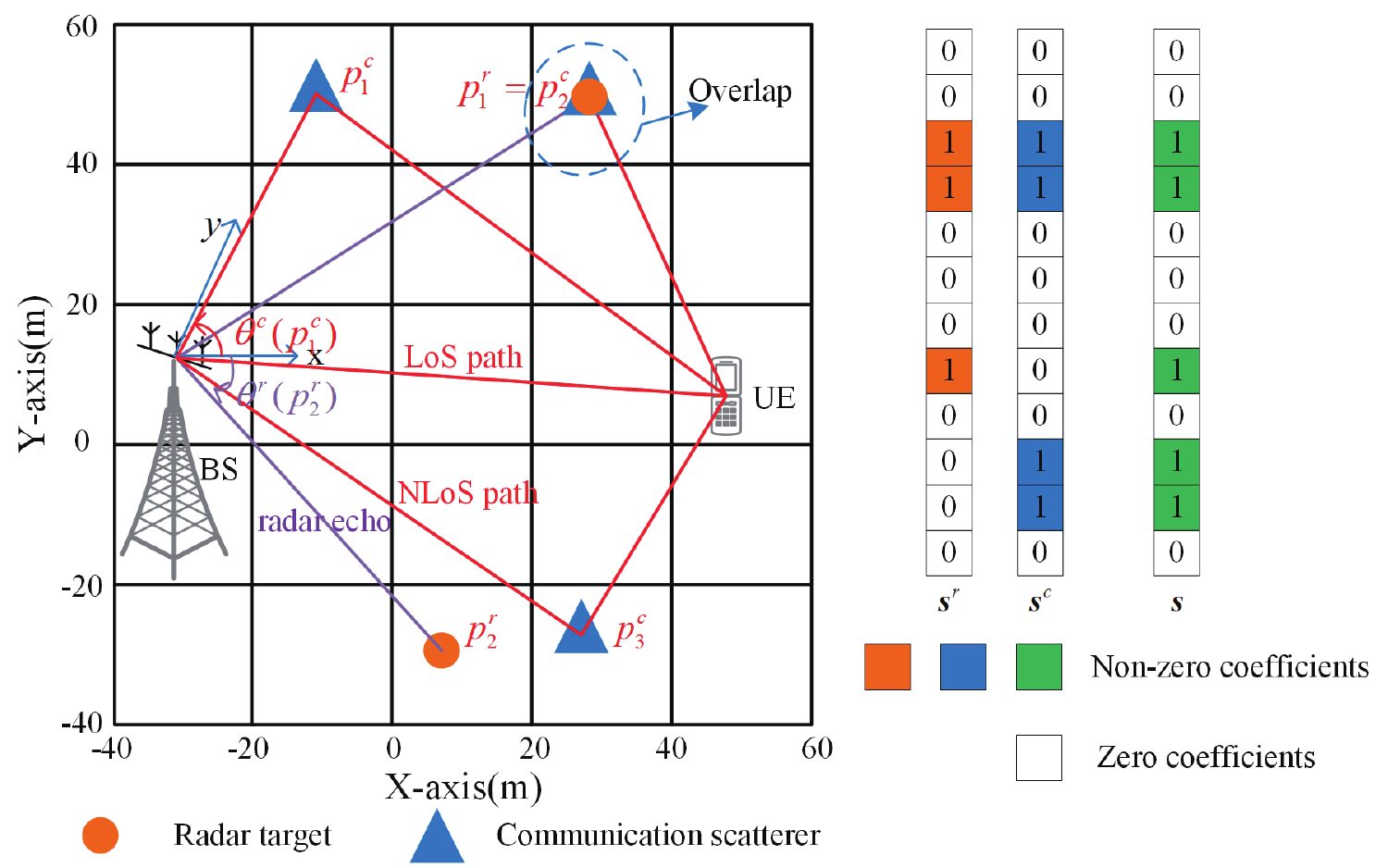}
\par\end{centering}
\caption{\label{fig:Illustration-of-radar}Illustration of radar and communication
channels.}
\end{figure}

Consider a TDD massive MIMO-OFDM ISAC system, where one BS equipped
with $M\gg1$ antennas serves a mobile user equipped with one antenna
while sensing the scattering environment, as illustrated in Fig. \ref{fig:Illustration-of-radar}.
The BS transmits downlink pilots to sense the targets in the scattering
environment, and then the user transmits uplink pilots to estimate
the communication channel. Suppose there are a total number of $K$
targets and $L$ communication scatterers in the scattering environment.
As discussed above, there might be some overlap between targets and
communication scatterers. The user is located at $\boldsymbol{p}_{u}=\left[p^{x},p^{y}\right]^{T}$
in a two-dimensional (2-D) area $\mathcal{R}$. The BS is located
at a known position $\boldsymbol{p}_{b}=\left[\tilde{p}^{x},\tilde{p}^{y}\right]^{T}$.
Let $\boldsymbol{p}_{\mathrm{\mathit{k}}}^{r}=\left[p_{k}^{r,x},p_{k}^{r,y}\right]^{T}$
and $\boldsymbol{p}_{\mathrm{\mathit{l}}}^{c}=\left[p_{l}^{c,x},p_{l}^{c,y}\right]^{T}$
be the coordinates of the $k\textrm{-th}$ target and the $l\textrm{-th}$
communication scatterer. We assume that the BS has some prior information
about the user location based on the Global Positioning System (GPS)
or the previous user localization result.

\subsection{Reflected Downlink Pilot Signal}

Target sensing aims at detecting the presence of the target and estimating
the target location. To achieve this, on the $n\textrm{-th}$ subcarrier,
the BS transmits a downlink pilot $\boldsymbol{v}_{n}^{r}\in\mathbb{C}^{M\times1}$,
and the received signal reflected from the targets can be expressed
as 
\begin{equation}
\boldsymbol{y}_{n}^{r}=\mathbf{H}_{n}^{r}\boldsymbol{v}_{n}^{r}+\boldsymbol{\boldsymbol{z}}_{n}^{r},
\end{equation}
where $\mathbf{H}_{n}^{r}\in\mathbb{C}^{M\times M}$ denotes the radar
channel matrix and $\boldsymbol{\boldsymbol{\boldsymbol{z}}}_{n}^{r}$
is the additive white Gaussian noise (AWGN) with variance $\left(\sigma_{z}^{r}\right)^{2}$.
Let $\theta^{r}\left(\boldsymbol{p}_{k}^{r}\right)$ and $\tau^{r}\left(\boldsymbol{p}_{k}^{r}\right)$
represent the angle of arrival (AoA) and delay of the $k\textrm{-th}$
target, respectively, which are related to the position of the BS
and the $k\textrm{-th}$ target through

\begin{align*}
\theta^{r}\left(\boldsymbol{p}_{k}^{r}\right) & =\arctan\left(\frac{p_{k}^{r,y}-\tilde{p}^{y}}{p_{k}^{r,x}-\tilde{p}^{x}}\right)+\pi\cdotp\mathds{1}\left(p_{k}^{r,x}<\tilde{p}^{x}\right),\\
\tau^{r}\left(\boldsymbol{p}_{k}^{r}\right) & =\frac{2\left\Vert \boldsymbol{p}_{b}-\boldsymbol{p}_{k}^{r}\right\Vert }{c},
\end{align*}
where the angle is calculated anticlockwise and with respect to the
\textit{x}-axis, $\mathds{1}\left(E\right)$ is the indication function,
which means that if the logical expression $E$ is true, then $\mathds{1}\left(E\right)=1$,
$\left\Vert \cdot\right\Vert $ denotes the Euclidean norm of the
given vector, and $c$ denotes the speed of light. Then the radar
channel matrix can be modeled as
\begin{equation}
\mathbf{H}_{n}^{r}=\sum_{k=0}^{K}x_{k}^{r}e^{-j2\pi nf_{0}\left(\tau^{r}\left(\boldsymbol{p}_{k}^{r}\right)\right)}\boldsymbol{a}\left(\theta^{r}\left(\boldsymbol{p}_{k}^{r}\right)\right)\boldsymbol{a}^{T}\left(\theta^{r}\left(\boldsymbol{p}_{k}^{r}\right)\right),\label{eq:Hnr}
\end{equation}
where $x_{k}^{r}$ represents radar cross section of the \textit{k}-th
target, $f_{0}$ is the subcarrier interval, and $\boldsymbol{a}\left(\theta\right)\in\mathbb{C}^{M\times1}$
denotes the array response vector at the BS. For the special case
of a uniform linear array (ULA), we have

\[
\boldsymbol{a}\left(\theta\right)=\frac{1}{\sqrt{M}}\left[1,e^{j\pi\sin\theta},\ldots,e^{j\left(M-1\right)\pi\sin\theta}\right]^{T}.
\]
Note that in (\ref{eq:Hnr}), we treat the user as the $0\textrm{-th}$
target and define $\boldsymbol{p}_{0}^{r}\triangleq\boldsymbol{p}_{u}$.
If the BS can ``see'' the user through the radar echo signal, we
have $x_{0}^{r}>0$. In this case, the echo signal also directly provides
some additional information to assist in locating the user's position.

\subsection{Received Uplink Pilot Signal}

On the \textit{n}-th subcarrier, the user transmits an uplink pilot
$u_{n}^{c}\in\mathbb{C}$ and then the BS receives the signal, which
can be expressed as 
\begin{equation}
\boldsymbol{y}_{n}^{c}=\boldsymbol{h}_{n}^{c}u_{n}^{c}+\boldsymbol{z}_{n}^{c},
\end{equation}
where $\boldsymbol{h}_{n}^{c}\in\mathbb{C}^{M\times1}$ denotes the
communication channel vector, $\boldsymbol{z}_{n}^{c}$ is the AWGN
with variance $\left(\sigma_{z}^{c}\right)^{2}$. 

Assume that there are $(L+1)$ paths for the communication channel,
i.e., one LoS path and $L$ non-LoS (NLoS) paths. For convenience,
we define the LoS path as the $0\textrm{-th}$ channel path and treat
the user as the $0\textrm{-th}$ communication scatterer with its
position $\boldsymbol{p}_{0}^{c}\triangleq\boldsymbol{p}_{u}$. Let
$\theta^{c}\left(\boldsymbol{p}_{l}^{c}\right)$ and $\tau^{c}\left(\boldsymbol{p}_{l}^{c},\boldsymbol{p}_{u}\right)$
represent the AoA and relative dealy of the $l\textrm{-th}$ channel
path, respectively, which are related to the position of the BS, the
user, and the $l\textrm{-th}$ communication scatterer through

\begin{align*}
\theta^{c}\left(\boldsymbol{p}_{l}^{c}\right) & =\arctan\left(\frac{p_{l}^{c,y}-\tilde{p}^{y}}{p_{l}^{c,x}-\tilde{p}^{x}}\right)+\pi\cdotp\mathds{1}\left(p_{l}^{c,x}<\tilde{p}^{x}\right),\\
\tau^{c}\left(\boldsymbol{p}_{l}^{c},\boldsymbol{p}_{u}\right) & =\frac{\left\Vert \boldsymbol{p}_{b}-\boldsymbol{p}_{l}^{c}\right\Vert }{c}+\frac{\left\Vert \boldsymbol{p}_{u}-\boldsymbol{p}_{l}^{c}\right\Vert }{c}-\frac{\left\Vert \boldsymbol{p}_{b}-\boldsymbol{p}_{u}\right\Vert }{c}.
\end{align*}
Then the communication channel vector can be expressed as 

\begin{equation}
\boldsymbol{h}_{n}^{c}=\sum_{l=0}^{L}x_{l}^{c}e^{-j2\pi nf_{0}\left(\tau^{c}\left(\boldsymbol{p}_{l}^{c},\boldsymbol{p}_{u}\right)+\tau_{o}\right)}\boldsymbol{a}\left(\theta^{c}\left(\boldsymbol{p}_{l}^{c}\right)\right),\label{eq:hnc}
\end{equation}
where $x_{l}^{c}$ denotes the complex gain of the $l\textrm{-th}$
channel path and $\tau_{o}$ is the time offset caused by the timing
synchronization error at the BS. 

The non-ideal factors of time offset and user localization error will
cause localization ambiguity for the communication scatterers and
degrade the performance of communication channel estimation. We will
elaborate on how to estimate these non-ideal factors based on the
EM method in Section \ref{sec:SEA-Turbo-SBI-Algorithm}. 

\section{Sparse Bayesian Inference Formulation }

In this section, we first obtain a sparse representation of the radar
and communication channels in the location domain. Then, we introduce
a sparse prior model to capture the partially common sparsity of the
radar and communication channels. Finally, we formulate the joint
scattering environment sensing and channel estimation problem as a
sparse Bayesian inference problem.

\subsection{A Location Domain Sparse Representation of Channels}

It is difficult to directly estimate the position of targets and communication
scatterers through maximum a posteriori (MAP) method because the optimization
problem is non-convex and has a lot of local optima. To solve this
issue, we introduce a grid-based solution to obtain a sparse representation
of the channels for better sensing and estimation performance. Specifically,
we define a 2-D uniform grid $\left\{ \overline{\boldsymbol{r}}_{1},\ldots,\overline{\boldsymbol{r}}_{Q}\right\} \subset\mathcal{R}$
of $Q\gg K+L$ positions, as illustrated in Fig. \ref{fig:Illustration-of-radar}.

In practice, the true positions usually do not lie exactly on the
$Q$ discrete position grid points. To get around this problem, one
common solution is to introduce a dynamic position grid, denoted by
$\boldsymbol{r}=\left[\boldsymbol{r}_{1};\ldots;\boldsymbol{r}_{Q}\right]$,
instead of only using a fixed position grid. In this case, there always
exists an $\boldsymbol{r}^{*}$ that covers the true position of all
targets and communication scatterers. In general, the uniform grid
is chosen as the initial point for $\boldsymbol{r}$ in the algorithm,
which makes it easier to find a near-optimal solution for the MAP
estimation problem. 

Then we define the sparse basis with a dynamic position grid for the
radar and communication channels as 

\[
\mathbf{A}\left(\boldsymbol{r},\boldsymbol{p}_{u}\right)\triangleq\left[\boldsymbol{a}\left(\theta^{r}\left(\boldsymbol{p}_{u}\right)\right),\widetilde{\mathbf{A}}\left(\boldsymbol{r}\right)\right]\in\mathbb{C}^{M\times\left(Q+1\right)},
\]
where

\[
\widetilde{\mathbf{A}}\left(\boldsymbol{r}\right)\triangleq\left[\boldsymbol{a}\left(\theta^{r}\left(\boldsymbol{r}_{1}\right)\right),\ldots,\boldsymbol{a}\left(\theta^{r}\left(\boldsymbol{r}_{Q}\right)\right)\right].
\]
The sparse representation of the radar channel matrix and the communication
channel vector on the $n\textrm{-th}$ subcarrier corresponding to
(\ref{eq:Hnr}) and (\ref{eq:hnc}) are respectively given by
\begin{align}
\mathbf{H}_{n}^{r} & =\mathbf{A}\left(\boldsymbol{r},\boldsymbol{p}_{u}\right)\mathbf{D}_{n}^{r}\textrm{diag}\left(\boldsymbol{x}^{r}\right)\mathbf{A}^{T}\left(\boldsymbol{r},\boldsymbol{p}_{u}\right),\label{eq:Hnr2}\\
\boldsymbol{h}_{n}^{c} & =\mathbf{A}\left(\boldsymbol{r},\boldsymbol{p}_{u}\right)\mathbf{D}_{n}^{c}\boldsymbol{x}^{c},\label{eq:hncNL2}
\end{align}
where $\boldsymbol{x}^{r}\in\mathbb{C}^{\left(Q+1\right)\times1}$
and $\boldsymbol{x}^{c}\in\mathbb{C}^{\left(Q+1\right)\times1}$ are
called the location domain sparse radar and communication channel
vectors, $\mathbf{D}_{n}^{r}$ and $\mathbf{D}_{n}^{c}$ are diagonal
matrices, with the $0\textrm{-th}$ diagonal elements being $e^{-j2\pi nf_{0}\tau^{r}\left(\boldsymbol{p}_{u}\right)}$
and $e^{-j2\pi nf_{0}\tau_{o}}$, respectively, and the $q\textrm{-th}$
diagonal elements being $e^{-j2\pi nf_{0}\tau^{r}\left(\boldsymbol{r}_{q}\right)}$
and $e^{-j2\pi nf_{0}\left(\tau^{c}\left(\boldsymbol{r}_{q},\boldsymbol{p}_{u}\right)+\tau_{o}\right)}$,
respectively, for $q=1,\ldots,Q$. $\boldsymbol{x}^{r}$ and $\boldsymbol{x}^{c}$
only have a few non-zero elements corresponding to the position of
targets and communication scatterers, respectively. Specifically,
the $q\textrm{-th}$ element of $\boldsymbol{x}^{r}$, denoted by
$x_{q}^{r}$, represents the complex reflection coefficient of a target
lying in the position $\boldsymbol{r}_{q}$. The $q\textrm{-th}$
element of $\boldsymbol{x}^{c}$, denoted by $x_{q}^{c}$, represents
the complex channel gain of the channel path with the corresponding
communication scatterer lying in the position $\boldsymbol{r}_{q}$.

\subsection{A Sparse Prior Model for the Partially Common Sparsity}

We introduce a sparse prior model to describe the partially common
sparsity of the location domain radar and communication channels.
We define the support vectors of the radar channel and communication
channel as $\boldsymbol{s}^{r}\triangleq\left[s_{0}^{r},\ldots,s_{Q}^{r}\right]^{T}$
and $\boldsymbol{s}^{c}\triangleq\left[s_{0}^{c},\ldots,s_{Q}^{c}\right]^{T}$,
respectively. If there is a radar target (communication scatterer)
around the $q\textrm{-th}$ position grid $\boldsymbol{r}_{q}$, we
have $s_{q}^{r}=1$ ($s_{q}^{c}=1$). Otherwise, we have $s_{q}^{r}=0$
($s_{q}^{c}=0$). Note that $s_{0}^{r}=1$ indicates that the BS can
``see'' the mobile user through the radar echo signal and $s_{0}^{c}=1$
indicates that the LoS path exists.

The elements of $\boldsymbol{x}^{r}$ and $\boldsymbol{x}^{c}$ are
independent conditioned on the support vectors $\boldsymbol{s}^{r}$
and $\boldsymbol{s}^{c}$, and the conditional distributions are given
by
\begin{subequations}
\begin{align}
p\left(x_{q}^{r}\mid s_{q}^{r}\right) & =\left(1-s_{q}^{r}\right)\delta\left(x_{q}^{r}\right)+s_{q}^{r}\mathcal{CN}\left(x_{q}^{r};0,\left(\sigma_{q}^{r}\right)^{2}\right),\label{eq:p(xqr|sqr)}\\
p\left(x_{q}^{c}\mid s_{q}^{c}\right) & =\left(1-s_{q}^{c}\right)\delta\left(x_{q}^{c}\right)+s_{q}^{c}\mathcal{CN}\left(x_{q}^{c};0,\left(\sigma_{q}^{c}\right)^{2}\right),\label{eq:p(xqc|sqc)}
\end{align}
\end{subequations}
where $\delta\left(\cdot\right)$ is the Dirac Delta function, $\left(\sigma_{q}^{r}\right)^{2}$
and $\left(\sigma_{q}^{c}\right)^{2}$ denote the conditional variance
of $x_{q}^{r}$ and $x_{q}^{c}$ , respectively.

Then we introduce a joint support vector $\boldsymbol{s}\triangleq\left[s_{0},\ldots,s_{Q}\right]^{T}$
with $s_{q}=s_{q}^{r}\vee s_{q}^{c}$ to represent the common positions
of the radar targets and communication scatterers, where $\vee$ means
the logical ``or'' operator. The joint distribution of support vectors
$\boldsymbol{s}^{r}$, $\boldsymbol{s}^{c}$ and $\boldsymbol{s}$
can be expressed as
\begin{align}
p\left(\boldsymbol{s}^{r},\boldsymbol{s}^{c},\boldsymbol{s}\right) & =p\left(\boldsymbol{s}^{r}\mid\boldsymbol{s}\right)p\left(\boldsymbol{s}^{c}\mid\boldsymbol{s}\right)p\left(\boldsymbol{s}\right)\nonumber \\
 & =\prod_{q}p\left(s_{q}^{r}\mid s_{q}\right)\prod_{q}p\left(s_{q}^{c}\mid s_{q}\right)\prod_{q}p\left(s_{q}\right),
\end{align}
where
\begin{subequations}
\begin{align}
p\left(s_{q}^{r}\mid s_{q}\right) & =\left(1-s_{q}\right)\delta\left(s_{q}^{r}\right)+s_{q}\left(\rho_{r}^{s_{q}^{r}}\left(1-\rho_{r}\right)^{1-s_{q}^{r}}\right),\label{eq:p(sqr|sq)}\\
p\left(s_{q}^{c}\mid s_{q}\right) & =\left(1-s_{q}\right)\delta\left(s_{q}^{c}\right)+s_{q}\left(\rho_{c}^{s_{q}^{c}}\left(1-\rho_{c}\right)^{1-s_{q}^{c}}\right),\label{eq:p(sqc|sq)}\\
p\left(s_{q}\right) & =\lambda^{s_{q}}\left(1-\lambda\right)^{1-s_{q}},\label{eq:p(sq)}
\end{align}
\end{subequations}
where $\lambda$ denotes the sparsity level of $\boldsymbol{s}$,
$\rho_{r}$ and $\rho_{c}$ represent the probability of $s_{q}^{r}=1$
and $s_{q}^{c}=1$ conditioned on $s_{q}=1$, respectively, and the
value of $\left(\rho_{r}+\rho_{c}-1\right)$ represents how much the
targets and communication scatterers overlap.

With the sparse prior model discussed above, the joint distribution
of all random variables can be expressed as

\begin{align}
 & p\left(\boldsymbol{s}^{r},\boldsymbol{s}^{c},\boldsymbol{s},\boldsymbol{x}^{r},\boldsymbol{x}^{c}\right)\nonumber \\
= & p\left(\boldsymbol{s}^{r},\boldsymbol{s}^{c},\boldsymbol{s}\right)\prod_{q}p\left(x_{q}^{r}\mid s_{q}^{r}\right)\prod_{q}p\left(x_{q}^{c}\mid s_{q}^{c}\right).
\end{align}

\subsection{Sparse Bayesian Inference with Uncertain Parameters}

Using the location domain sparse representation of radar channel and
communication channel in (\ref{eq:Hnr2}) and (\ref{eq:hncNL2}),
the reflected downlink pilot signal and received uplink pilot signal
on all available subcarriers can be expressed as
\begin{subequations}
\begin{align}
\boldsymbol{y}^{r} & =\boldsymbol{\Phi}^{r}\left(\boldsymbol{r},\boldsymbol{p}_{u}\right)\boldsymbol{x}^{r}+\boldsymbol{z}^{r},\label{eq:yr}\\
\boldsymbol{y}^{c} & =\boldsymbol{\Phi}^{c}\left(\boldsymbol{r},\boldsymbol{p}_{u},\tau_{o}\right)\boldsymbol{x}^{c}+\boldsymbol{z}^{c},\label{eq:yc}
\end{align}
\end{subequations}
where $\boldsymbol{y}^{r}$, $\boldsymbol{y}^{c}$, $\boldsymbol{z}^{r}$,
and $\boldsymbol{z}^{c}$ are respectively given by

\begin{align*}
\boldsymbol{y}^{r} & \triangleq\left[\left(\boldsymbol{y}_{1}^{r}\right)^{T},\ldots,\left(\boldsymbol{y}_{N}^{r}\right)^{T}\right]^{T}\in\mathbb{C}^{MN\times1},\\
\boldsymbol{y}^{c} & \triangleq\left[\left(\boldsymbol{y}_{1}^{c}\right)^{T},\ldots,\left(\boldsymbol{y}_{N}^{c}\right)^{T}\right]^{T}\in\mathbb{C}^{MN\times1},\\
\boldsymbol{z}^{r} & \triangleq\left[\left(\boldsymbol{z}_{1}^{r}\right)^{T},\ldots,\left(\boldsymbol{z}_{N}^{r}\right)^{T}\right]^{T}\in\mathbb{C}^{MN\times1},\\
\boldsymbol{z}^{c} & \triangleq\left[\left(\boldsymbol{z}_{1}^{c}\right)^{T},\ldots,\left(\boldsymbol{z}_{N}^{c}\right)^{T}\right]^{T}\in\mathbb{C}^{MN\times1},
\end{align*}
$\boldsymbol{\Phi}^{c}\in\mathbb{C}^{MN\times\left(Q+1\right)}$ denotes
the communication measurement matrix, which is given by
\[
\boldsymbol{\Phi}^{c}=\left[\begin{array}{c}
u_{1}^{c}\mathbf{A}\left(\boldsymbol{r},\boldsymbol{p}_{u}\right)\mathbf{D}_{1}^{c}\\
\cdots\\
u_{N}^{c}\mathbf{A}\left(\boldsymbol{r},\boldsymbol{p}_{u}\right)\mathbf{D}_{N}^{c}
\end{array}\right],
\]
and $\boldsymbol{\Phi}^{r}\in\mathbb{C}^{MN\times\left(Q+1\right)}$
denotes the radar measurement matrix that consists of the $\left(\left(q-1\right)Q+q\right)\textrm{-th}$
column of $\widetilde{\boldsymbol{\Phi}}^{r}\in\mathbb{C}^{MN\times\left(Q+1\right)^{2}}$
for $q=1,\ldots,\left(Q+1\right)$, where

\[
\widetilde{\boldsymbol{\Phi}}^{r}=\left[\begin{array}{c}
\left(\left(v_{1}^{r}\right)^{T}\mathbf{A}\left(\boldsymbol{r},\boldsymbol{p}_{u}\right)\right)\otimes\left(\mathbf{A}\left(\boldsymbol{r},\boldsymbol{p}_{u}\right)\mathbf{D}_{1}^{r}\right)\\
\cdots\\
\left(\left(v_{N}^{r}\right)^{T}\mathbf{A}\left(\boldsymbol{r},\boldsymbol{p}_{u}\right)\right)\otimes\left(\mathbf{A}\left(\boldsymbol{r},\boldsymbol{p}_{u}\right)\mathbf{D}_{N}^{r}\right)
\end{array}\right],
\]
where $\otimes$ means the Kronecker product operator. For convenience,
we combine (\ref{eq:yr}) and (\ref{eq:yc}) into a linear observation
model as

\begin{equation}
\boldsymbol{y}=\boldsymbol{\Phi}\left(\boldsymbol{\xi}\right)\boldsymbol{x}+\boldsymbol{z},\label{eq:y}
\end{equation}
where $\boldsymbol{\xi}\triangleq\left\{ \boldsymbol{r},\boldsymbol{p}_{u},\tau_{o}\right\} $
is the collection of sensing parameters, $\boldsymbol{y}\triangleq\left[\left(\boldsymbol{y}^{r}\right)^{T},\left(\boldsymbol{y}^{c}\right)^{T}\right]^{T}$,
$\boldsymbol{x}\triangleq\left[\left(\boldsymbol{x}^{r}\right)^{T},\left(\boldsymbol{x}^{c}\right)^{T}\right]^{T}$,
$\boldsymbol{z}\triangleq\left[\left(\boldsymbol{z}^{r}\right)^{T},\left(\boldsymbol{z}^{c}\right)^{T}\right]^{T}$,
and $\boldsymbol{\Phi}\left(\boldsymbol{\xi}\right)\triangleq\mathrm{\textrm{BlockDiag}}\left(\boldsymbol{\Phi}^{r},\boldsymbol{\Phi}^{c}\right)$.

Let $p\left(\boldsymbol{\xi}\right)$ represents the known prior distribution
of the sensing parameters (we can assume uniform distribution if unknown).
Our primary goal is to estimate the channel vector $\boldsymbol{x}$,
the support set $\left\{ \boldsymbol{s}^{r},\boldsymbol{s}^{c}\right\} $,
and the uncertain parameters $\boldsymbol{\xi}$ given observation
$\boldsymbol{y}$ in model (\ref{eq:y}). To be specific, for given
$\boldsymbol{\xi}$, we aim at computing the conditional marginal
posteriors, i.e., $p\left(x_{q}^{r}\mid\boldsymbol{y};\boldsymbol{\xi}\right)$,
$p\left(x_{q}^{c}\mid\boldsymbol{y};\boldsymbol{\xi}\right)$, $p\left(s_{q}^{r}\mid\boldsymbol{y};\boldsymbol{\xi}\right)$,
$p\left(s_{q}^{c}\mid\boldsymbol{y};\boldsymbol{\xi}\right)$, $\forall q$.
On the other hand, the uncertain parameters $\boldsymbol{\xi}$ are
obtained by the MAP estimator as follows: 
\begin{align}
\boldsymbol{\xi}^{\ast} & =\underset{\boldsymbol{\xi}}{\arg\max}\ln p\left(\boldsymbol{y},\boldsymbol{\xi}\right).\label{eq:M-step}
\end{align}
Once we obtain the MAP estimate of $\boldsymbol{\xi}^{\ast}$, we
can obtain the minimum mean square error (MMSE) estimate of $x_{q}^{r}$
as $x_{q}^{r\ast}=\int_{x_{q}^{r}}x_{q}^{r}p\left(x_{q}^{r}\mid\boldsymbol{y};\boldsymbol{\xi}^{\ast}\right)$
and the MAP estimate of $s_{q}^{r}$ as $s_{q}^{r\ast}=\arg\max_{s_{q}^{r}}p\left(s_{q}^{r}\mid\boldsymbol{y};\boldsymbol{\xi}^{\ast}\right)$.
The MMSE estimate of $x_{q}^{c}$ and the MAP estimate of $s_{q}^{c}$
can be obtained in the same way.

However, the corresponding factor graph of the probability model contains
loops. Therefore, it is exceedingly challenging to calculate the above
conditional marginal posteriors precisely. In the following section,
we present the SEA-Turbo-SBI algorithm, which uses the turbo approach
to calculate approximate marginal posteriors and applies the EM method
to find an approximate solution for (\ref{eq:M-step}).

\section{SEA-Turbo-SBI Algorithm\label{sec:SEA-Turbo-SBI-Algorithm} }

The primary goal of the SEA-Turbo-SBI algorithm is to simultaneously
maximize $\ln p\left(\boldsymbol{y},\boldsymbol{\xi}\right)$ with
respect to the uncertain parameters $\boldsymbol{\xi}$ in (\ref{eq:M-step})
and approximately calculate the conditional marginal posteriors. The
SEA-Turbo-SBI algorithm, which is based on the EM method, iterates
between the next two major steps until convergence.
\begin{itemize}
\item \textbf{SEA-Turbo-SBI-E Step:} Based on the turbo approach, calculate
the approximate marginal posteriors, i.e., $p\left(x_{q}^{r}\mid\boldsymbol{y};\boldsymbol{\xi}^{i}\right)$,
$p\left(x_{q}^{c}\mid\boldsymbol{y};\boldsymbol{\xi}^{i}\right)$,
$p\left(s_{q}^{r}\mid\boldsymbol{y};\boldsymbol{\xi}^{i}\right)$,
$p\left(s_{q}^{c}\mid\boldsymbol{y};\boldsymbol{\xi}^{i}\right)$,
$\forall q$ for given $\boldsymbol{\xi}^{i}$ in the $i\textrm{-th}$
iteration.
\item \textbf{SEA-Turbo-SBI-M Step:} Construct a surrogate function for
$\ln p\left(\boldsymbol{y},\boldsymbol{\xi}\right)$ based on the
approximate marginal posterior $p\left(\boldsymbol{x}\mid\boldsymbol{y};\boldsymbol{\xi}^{i}\right)$
obtained in the SEA-Turbo-SBI-E Step, then use the gradient ascent
method to maximize the surrogate function with respect to $\boldsymbol{\xi}$.
\end{itemize}

\subsection{SEA-Turbo-SBI-E Step}

\begin{figure}[htbp]
\begin{centering}
\textsf{\includegraphics[width=80mm]{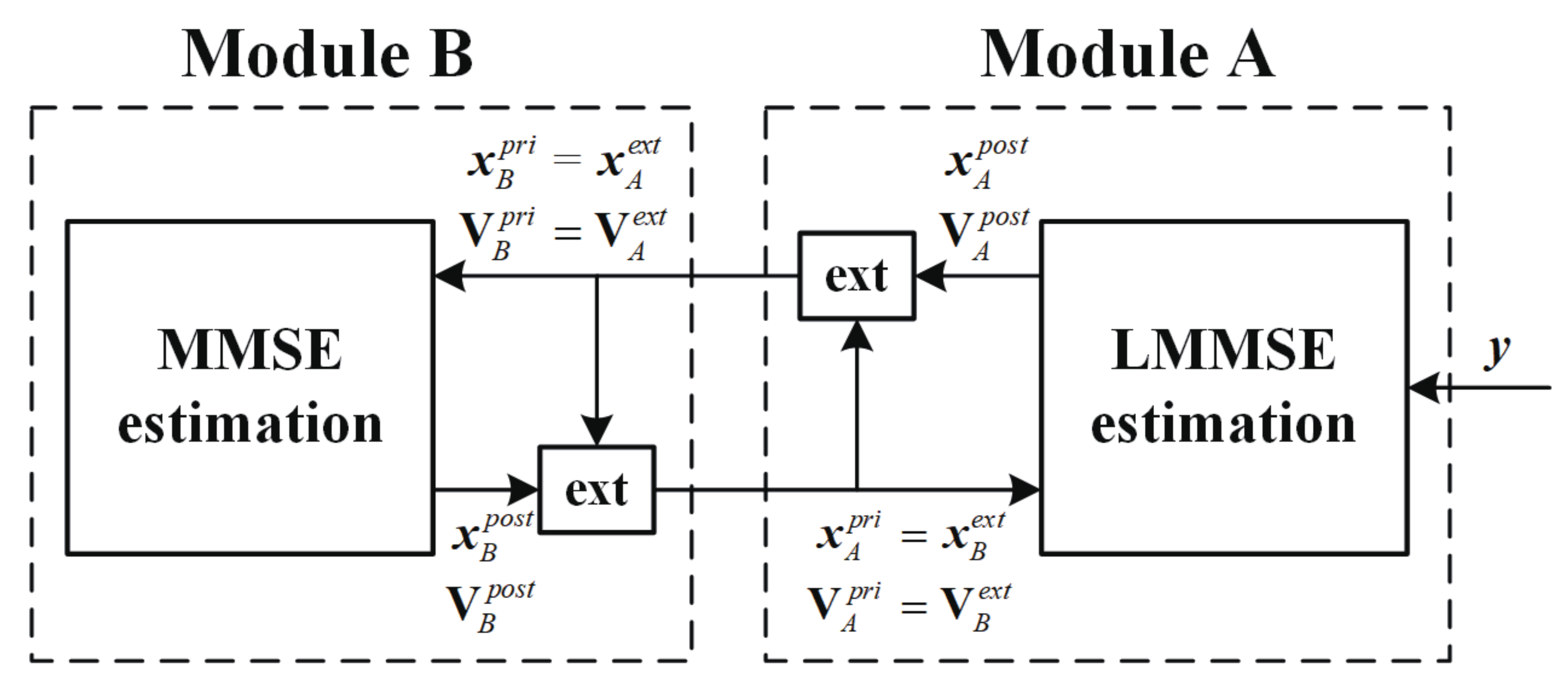}}
\par\end{centering}
\caption{\label{fig:Turbo-SBI}Illustration of the turbo approach.}
\end{figure}

There are two modules in the SEA-Turbo-SBI-E Step, as illustrated
in Fig. \ref{fig:Turbo-SBI}. Module A performs the linear minimum
mean square error (LMMSE) estimation based on the observation $\boldsymbol{y}$
and extrinsic messages from Module B, whereas Module B is a MMSE estimator
that can process the sparse prior information and extrinsic messages
from Module A. The two modules iterate until they reach a point of
convergence. We omit $\boldsymbol{\xi}$ in $\boldsymbol{\Phi}\left(\boldsymbol{\xi}\right)$
for simplicity in this subsection because $\boldsymbol{\xi}$ is fixed
in the SEA-Turbo-SBI-E Step.

\subsubsection{Module A with LMMSE Estimation}

We assume that the prior distribution of $\boldsymbol{x}$ is $\mathcal{CN}\left(\boldsymbol{x};\boldsymbol{x}_{A}^{pri},\mathbf{V}_{A}^{pri}\right)$,
where $\boldsymbol{x}_{A}^{pri}$ and $\mathbf{V}_{A}^{pri}$ are
the extrinsic mean and covariance matrix from Module B, respectively.
The posterior mean and covariance matrix of the LMMSE estimation are
respectively given by
\begin{align}
\boldsymbol{x}_{A}^{post} & =\mathbf{V}_{A}^{post}\left(\left(\mathbf{V}_{A}^{pri}\right)^{-1}\boldsymbol{x}_{A}^{pri}+\frac{\boldsymbol{\Phi}^{H}\boldsymbol{y}}{\sigma_{z}^{2}}\right),\label{eq:XApost}\\
\mathbf{V}_{A}^{post} & =\left(\frac{\boldsymbol{\Phi}^{H}\boldsymbol{\Phi}}{\sigma_{z}^{2}}+\left(\mathbf{V}_{A}^{pri}\right)^{-1}\right)^{-1}.\label{eq:VApost}
\end{align}
By subtracting the prior information from posterior information, we
obtain the extrinsic message from Module A as follows: 
\begin{align}
\boldsymbol{x}_{A}^{ext} & =\mathbf{V}_{A}^{ext}\left(\left(\overline{\mathbf{V}}_{A}^{post}\right)^{-1}\boldsymbol{x}_{A}^{post}-\left(\mathbf{V}_{A}^{pri}\right)^{-1}\boldsymbol{x}_{A}^{pri}\right),\nonumber \\
\mathbf{V}_{A}^{ext} & =\left(\left(\overline{\mathbf{V}}_{A}^{post}\right)^{-1}-\left(\mathbf{V}_{A}^{pri}\right)^{-1}\right)^{-1},
\end{align}
where $\overline{\mathbf{V}}_{A}^{post}$ takes each element in the
diagonal of matrix $\mathbf{V}_{A}^{post}$ while setting the non-diagonal
elements to be zero.

\subsubsection{Module B with Message Passing }

\begin{figure}[htbp]
\begin{centering}
\textsf{\includegraphics[width=80mm]{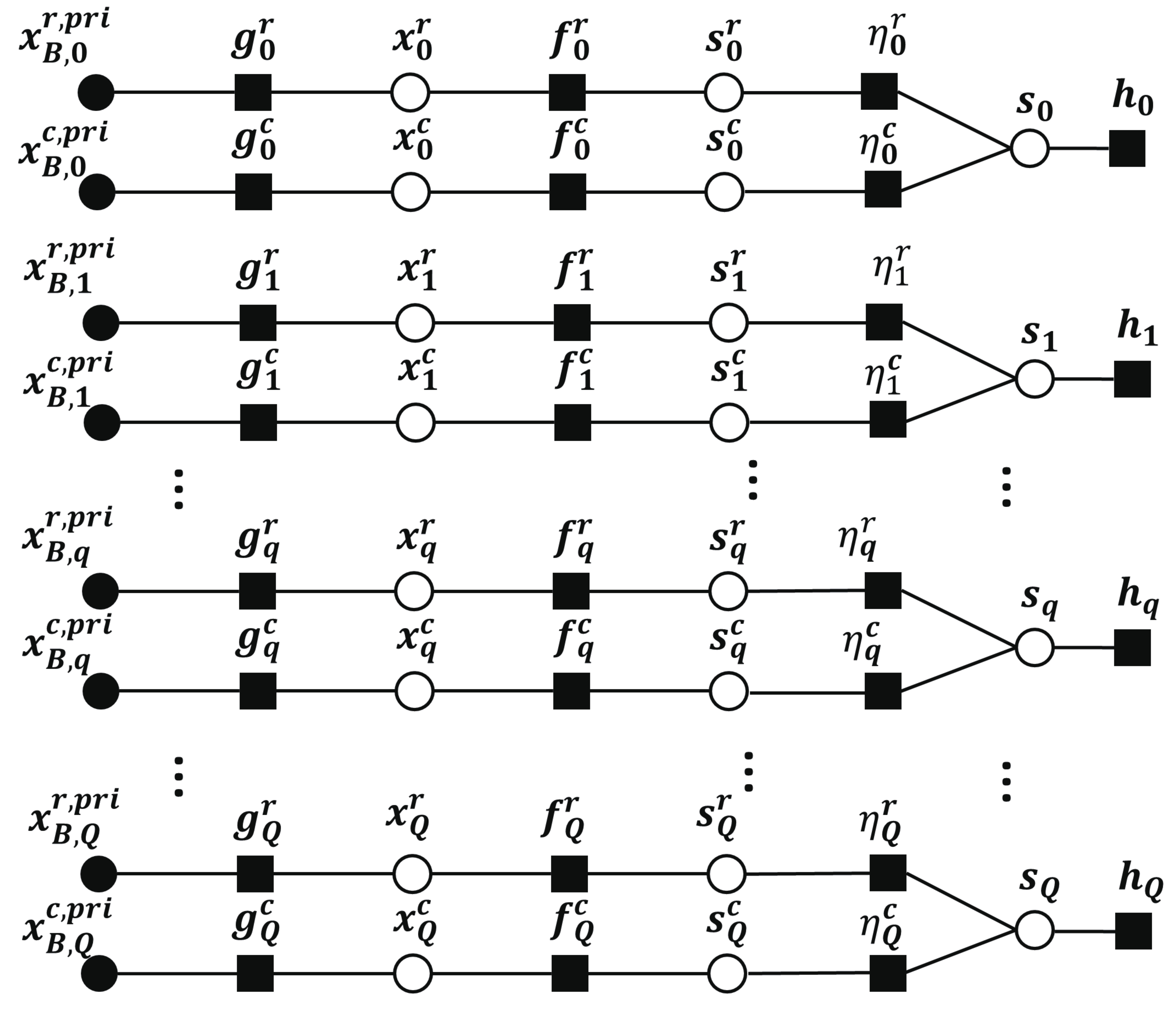}}
\par\end{centering}
\caption{\label{fig:Factor-graph}The factor graph of the joint distribution
of all variables.}
\end{figure}

In Module B, we construct a factor graph and derive a massage passing
algorithm to achieve the MMSE estimator. First of all, a basic assumption
is to model the extrinsic mean from Model A as an AWGN observation,
i.e.,
\begin{equation}
\boldsymbol{x}_{B}^{pri}=\boldsymbol{x}+\boldsymbol{w},
\end{equation}
where $\boldsymbol{w}\sim\mathcal{CN}\left(0,\mathbf{V}_{B}^{pri}\right)$
is the virtual noise, $\boldsymbol{x}_{B}^{pri}$ and $\mathbf{V}_{B}^{pri}$
are the extrinsic mean and covariance matrix from Module A, which
are respectively given by

\begin{align*}
\boldsymbol{x}_{B}^{pri} & =\boldsymbol{x}_{A}^{ext}\triangleq\left[\left(\boldsymbol{x}_{B}^{r,pri}\right)^{T},\left(\boldsymbol{x}_{B}^{c,pri}\right)^{T}\right]^{T},\\
\mathbf{V}_{B}^{pri} & =\mathbf{V}_{A}^{ext}\triangleq\textrm{BlockDiag}\left(\mathbf{V}_{B}^{r,pri},\mathbf{V}_{B}^{c,pri}\right).
\end{align*}
The factor graph of $p\left(\boldsymbol{x}_{B}^{r,pri},\boldsymbol{x}_{B}^{c,pri},\boldsymbol{x}^{r},\boldsymbol{x}^{c},\boldsymbol{s},\boldsymbol{s}^{r},\boldsymbol{s}^{c}\right)$,
denoted by $\mathcal{G}_{B}$, is shown in Fig. \ref{fig:Factor-graph},
where the factor nodes are defined as follows:

\begin{align*}
g_{q}^{t} & \triangleq\mathcal{CN}\left(x_{q}^{t};x_{B,q}^{t,pri},v_{B,q}^{t,pri}\right),t\in\left\{ r,c\right\} ,\forall q,\\
f_{q}^{t} & \triangleq p\left(x_{q}^{t}\mid s_{q}^{t}\right),t\in\left\{ r,c\right\} ,\forall q,\\
\eta_{q}^{t} & \triangleq p\left(s_{q}^{t}\mid s_{q}\right),t\in\left\{ r,c\right\} ,\forall q,\\
h_{q} & \triangleq p\left(s_{q}\right),\forall q,
\end{align*}
where $x_{B,q}^{t,pri}$ denotes the $q\textrm{-th}$ element of $\boldsymbol{x}_{B}^{t,pri}$
and $v_{B,q}^{t,pri}$ denotes the $q\textrm{-th}$ diagonal element
of $\mathbf{V}_{B}^{t,pri}$.

We use the sum-product rule to derive messages over the factor graph
$\mathcal{G}_{B}$ in Fig. \ref{fig:Factor-graph}. Due to the tree-type
structure of $\mathcal{G}_{B}$, the derivation of all messages is
relatively easy (similar to the message passing in Appendix A of \cite{Huangzhe_JRC2})
so that we omit it for simplicity. The approximate posterior distributions
can be calculated as
\begin{align}
\hat{p}\left(x_{q}^{t}\mid\boldsymbol{y}\right) & \propto\nu_{f_{q}^{t}\rightarrow x_{q}^{t}}\times\nu_{x_{q}^{t}\rightarrow f_{q}^{t}},t\in\left\{ r,c\right\} ,\forall q,\\
\hat{p}\left(s_{q}^{t}\mid\boldsymbol{y}\right) & \propto\nu_{f_{q}^{t}\rightarrow s_{q}^{t}}\times\nu_{s_{q}^{t}\rightarrow f_{q}^{t}},t\in\left\{ r,c\right\} ,\forall q,
\end{align}
where $\nu_{f_{q}^{t}\rightarrow x_{q}^{t}}$ and $\nu_{f_{q}^{t}\rightarrow s_{q}^{t}}$
denote the messages from factor node $f_{q}^{t}$ to variable nodes
$x_{q}^{t}$ and $s_{q}^{t}$, respectively, $\nu_{x_{q}^{t}\rightarrow f_{q}^{t}}$
and $\nu_{s_{q}^{t}\rightarrow f_{q}^{t}}$ denote the messages from
variable nodes $x_{q}^{t}$ and $s_{q}^{t}$ to factor node $f_{q}^{t}$,
respectively. Based on the the posterior distributions, the posterior
mean and covariance matrix for $\boldsymbol{x}^{t}$, denoted by

\begin{align*}
\boldsymbol{x}_{B}^{t,post} & \triangleq\left[x_{B,0}^{t,post},\ldots,x_{B,Q}^{t,post}\right]^{T},\\
\mathbf{V}_{B}^{t,post} & \triangleq\textrm{diag}\left(\left[v_{B,0}^{t,post},\ldots,v_{B,Q}^{t,post}\right]\right),
\end{align*}
can be calculated respectively as
\begin{align}
x_{B,q}^{t,post} & =\int_{x_{q}^{t}}x_{q}^{t}\hat{p}\left(x_{q}^{t}\mid\boldsymbol{y}\right),\\
v_{B,q}^{t,post} & =\int_{x_{q}^{t}}\left|x_{q}^{t}-x_{B,q}^{t,post}\right|^{2}\hat{p}\left(x_{q}^{t}\mid\boldsymbol{y}\right),
\end{align}
for $q=0,\ldots,Q$. Then the extrinsic message from Module B can
be calculated as
\begin{align}
\boldsymbol{x}_{B}^{ext} & =\mathbf{V}_{B}^{ext}\left(\left(\boldsymbol{V}_{B}^{post}\right)^{-1}\boldsymbol{x}_{B}^{post}-\left(\mathbf{V}_{A}^{pri}\right)^{-1}\boldsymbol{x}_{B}^{pri}\right),\nonumber \\
\mathbf{V}_{B}^{ext} & =\left(\left(\mathbf{V}_{B}^{post}\right)^{-1}-\left(\mathbf{V}_{B}^{pri}\right)^{-1}\right)^{-1},
\end{align}
where $\boldsymbol{x}_{B}^{post}$ and $\mathbf{V}_{B}^{post}$ are
respectively given by 
\begin{align*}
\boldsymbol{x}_{B}^{post} & \triangleq\left[\left(\boldsymbol{x}_{B}^{r,post}\right)^{T},\left(\boldsymbol{x}_{B}^{c,post}\right)^{T}\right]^{T},\\
\mathbf{V}_{B}^{post} & \triangleq\textrm{BlockDiag}\left(\mathbf{V}_{B}^{r,post},\mathbf{V}_{B}^{c,post}\right).
\end{align*}

\subsection{SEA-Turbo-SBI-M Step}

Since there is no close-form expression of $\ln p\left(\boldsymbol{y},\boldsymbol{\xi}\right)$,
it is challenging to directly solve the maximization problem in (\ref{eq:M-step}).
To get around this problem, one common solution is to construct a
surrogate function of $\ln p\left(\boldsymbol{y},\boldsymbol{\xi}\right)$
and maximize the surrogate function with respect to $\boldsymbol{\xi}$.
Specifically, in the $i\textrm{-th}$ iteration, the surrogate function
inspired by the EM method is given by 

\begin{align}
Q\left(\boldsymbol{\xi};\boldsymbol{\xi}^{i}\right)= & \int p\left(\boldsymbol{x}\mid\boldsymbol{y};\boldsymbol{\xi}^{i}\right)\ln\frac{p\left(\boldsymbol{y},\boldsymbol{x}\mid\boldsymbol{\xi}\right)}{p\left(\boldsymbol{x}\mid\boldsymbol{y};\boldsymbol{\xi}^{i}\right)}d\boldsymbol{x}+\ln p\left(\text{\ensuremath{\boldsymbol{\xi}}}\right)\nonumber \\
= & -\left(\sigma_{z}\right)^{-2}\left[\left\Vert \boldsymbol{y}-\boldsymbol{\Phi}\left(\boldsymbol{\xi}\right)\boldsymbol{x}^{post}\right\Vert ^{2}\right.\nonumber \\
 & +\left.\textrm{tr}\left(\boldsymbol{\Phi}\left(\boldsymbol{\xi}\right)\mathbf{V}^{post}\boldsymbol{\Phi}\left(\boldsymbol{\xi}\right)^{H}\right)\right]+\ln p\left(\text{\ensuremath{\boldsymbol{\xi}}}\right)+C,
\end{align}
where the posterior mean $\boldsymbol{x}^{post}$ and covariance matrix
$\mathbf{V}^{post}$ can be approximated to $\boldsymbol{x}_{A}^{post}$
and $\mathbf{V}_{A}^{post}$ in (\ref{eq:XApost}) and (\ref{eq:VApost}),
respectively, and $C$ is a constant. At the current iterate $\boldsymbol{\xi}^{i}$,
the surrogate function and its gradient satisfy the following properties:
\begin{subequations}
\begin{align}
Q\left(\boldsymbol{\xi};\boldsymbol{\xi}^{i}\right) & \leq\ln p\left(\boldsymbol{y},\boldsymbol{\xi}\right),\forall\boldsymbol{\xi},\\
Q\left(\boldsymbol{\xi}^{i};\boldsymbol{\xi}^{i}\right) & =\ln p\left(\boldsymbol{y},\boldsymbol{\xi}^{i}\right),\\
\frac{\partial Q\left(\boldsymbol{\xi};\boldsymbol{\xi}^{i}\right)}{\partial\boldsymbol{\xi}}\mid_{\boldsymbol{\xi}=\boldsymbol{\xi}^{i}} & =\frac{\partial\ln p\left(\boldsymbol{y},\boldsymbol{\xi}\right)}{\partial\boldsymbol{\xi}}\mid_{\boldsymbol{\xi}=\boldsymbol{\xi}^{i}}.
\end{align}
\end{subequations}
Then we need to maximize $Q\left(\boldsymbol{\xi};\boldsymbol{\xi}^{i}\right)$
to update the next iterate $\boldsymbol{\xi}^{i+1}$. However, it
is difficult to find the global optimal solution to the maximizing
problem because the function $Q\left(\boldsymbol{\xi};\boldsymbol{\xi}^{i}\right)$
is non-convex. Using the gradient ascent method, we can simply obtain
the next iterate $\boldsymbol{\xi}^{i+1}$ as
\begin{subequations}
\begin{align}
\boldsymbol{r}^{i+1} & =\boldsymbol{r}^{i}+\varepsilon_{r}^{i}\frac{\partial Q\left(\boldsymbol{r}^{i},\boldsymbol{p}_{u}^{i},\tau_{o}^{i};\boldsymbol{\xi}^{i}\right)}{\partial\boldsymbol{r}},\\
\boldsymbol{p}_{u}^{i+1} & =\boldsymbol{p}_{u}^{i}+\varepsilon_{p}^{i}\frac{\partial Q\left(\boldsymbol{r}^{i+1},\boldsymbol{p}_{u}^{i},\tau_{o}^{i};\boldsymbol{\xi}^{i}\right)}{\partial\boldsymbol{p}_{u}},\\
\tau_{o}^{i+1} & =\tau_{o}^{i}+\varepsilon_{t}^{i}\frac{\partial Q\left(\boldsymbol{r}^{i+1},\boldsymbol{p}_{u}^{i+1},\tau_{o}^{i};\boldsymbol{\xi}^{i}\right)}{\partial\tau_{o}},
\end{align}
\end{subequations}
where $\varepsilon_{r}^{i}$, $\varepsilon_{p}^{i}$ and $\varepsilon_{t}^{i}$
are step sizes determined by the Armijo rule. As a result, we have
$\ln p\left(\boldsymbol{y},\boldsymbol{\xi}^{i+1}\right)\geq Q\left(\boldsymbol{\xi}^{i+1};\boldsymbol{\xi}^{i}\right)\geq Q\left(\boldsymbol{\xi}^{i};\boldsymbol{\xi}^{i}\right)=\ln p\left(\boldsymbol{y},\boldsymbol{\xi}^{i}\right)$,
which indicates that the function $\ln p\left(\boldsymbol{y},\boldsymbol{\xi}\right)$
increases strictly until it reaches a stationary point.

\section{Simulation Results}

In the simulations, we consider a $100\text{ m}\times100\text{ m}$
area with a grid resolution of $5\text{ m}$. The BS is at coordinates
$\left[-50\text{ m},0\text{ m}\right]$ and the mobile user is around
coordinates $\left[50\text{ m},0\text{ m}\right]$ with a random position
offset. We assume that the prior information about the user location
is $p^{x}\sim\mathcal{N}\left(50,\sigma_{p}^{2}/2\right)$ and $p^{y}\sim\mathcal{N}\left(0,\sigma_{p}^{2}/2\right)$,
where $\sigma_{p}^{2}$ is set as $1$. There are $K=9$ radar targets
and $L=10$ communication scatterers within the area. The number of
OFDM subcarriers is $N=1024$ and the subcarrier interval is $f_{0}=30\text{ kHz}$.
Pilot symbols are inserted at intervals of $32$ OFDM subcarriers.
The BS is equipped with a ULA of $M=64$ antennas. The time offset
$\tau_{o}$ is within $\left[\frac{-2}{B},\frac{2}{B}\right]$, where
$B=Nf_{0}$ denotes the total bandwidth. We compare the performance
for orthogonal matching pursuit (OMP) based on separate estimation
\cite{Tropp_CE_OMP}, turbo compressed sensing (Turbo-CS) based on
joint estimation with a fixed position grid \cite{Yuan_TurboCS,LiuAn_CE_Turbo_CS},
the proposed SEA-Turbo-SBI based on separate estimation, i.e., without
the joint support vector $\boldsymbol{s}$, and the proposed SEA-Turbo-SBI
based on joint estimation.

\begin{figure}[tbh]
\begin{centering}
\includegraphics[width=68mm]{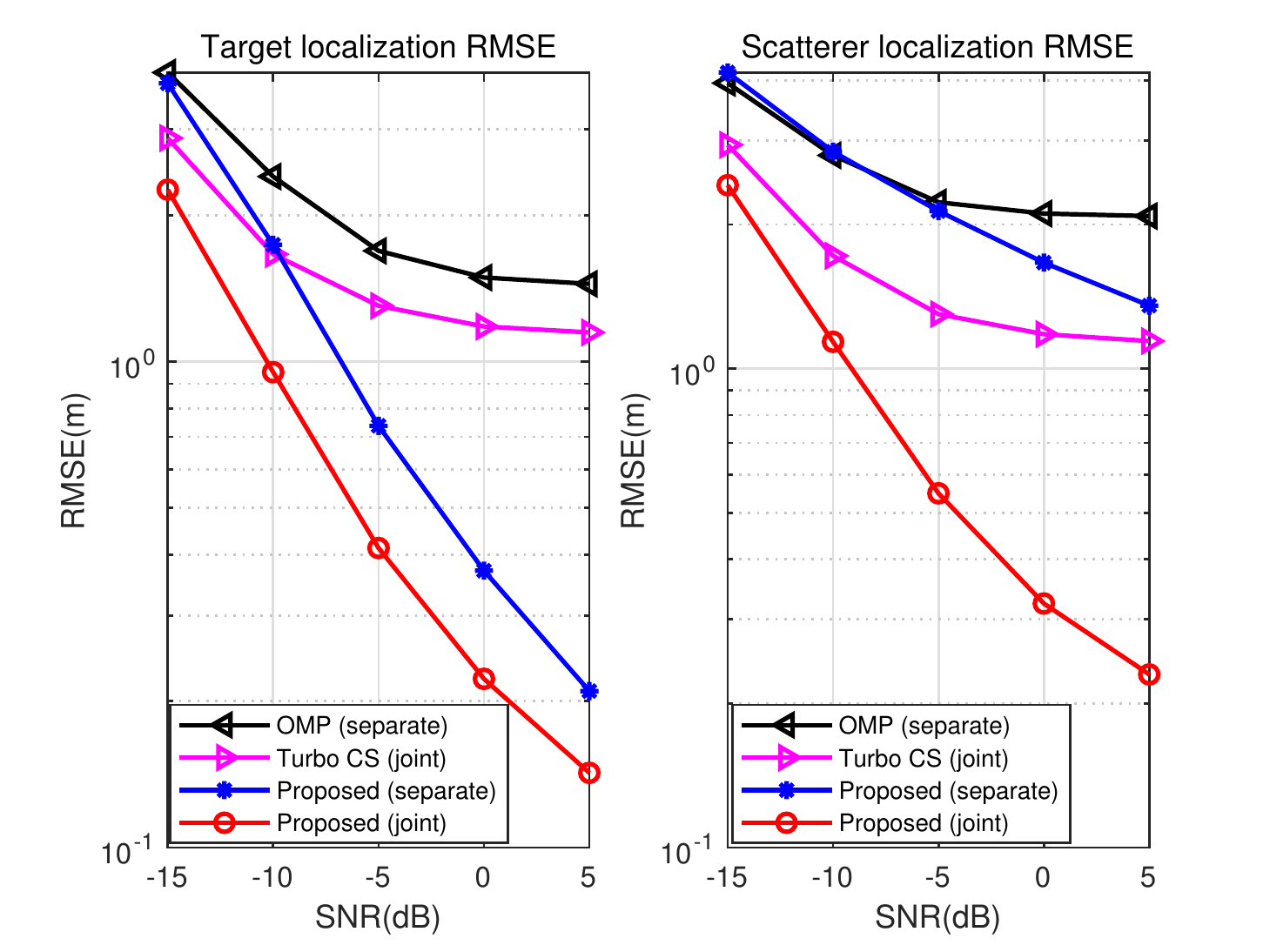}
\par\end{centering}
\caption{\label{fig:RMSE-versus-SNR} RMSE of target/scatterer localization
versus SNR.}
\end{figure}

\begin{figure}[tbh]
\begin{centering}
\includegraphics[width=68mm]{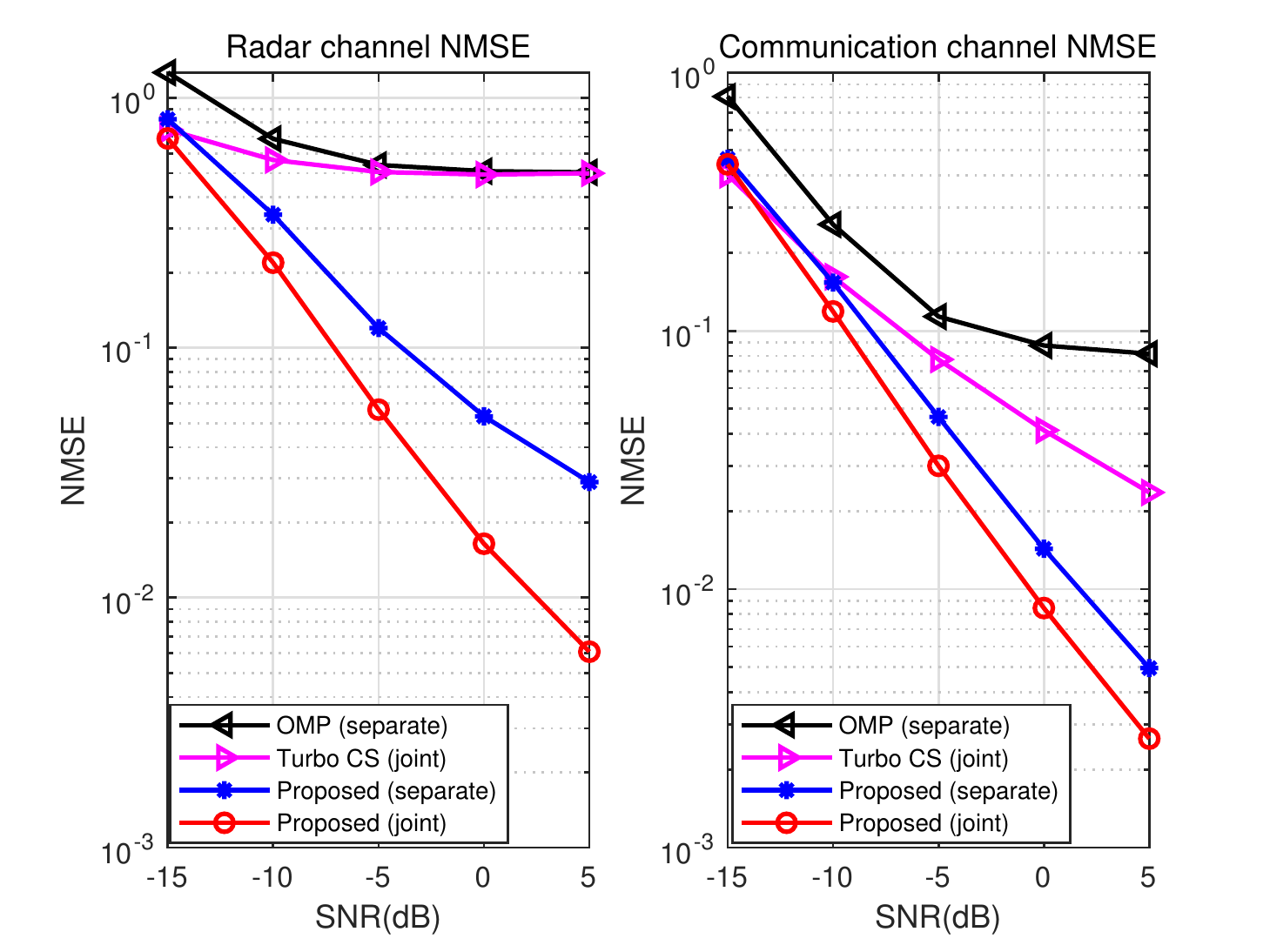}
\par\end{centering}
\caption{\label{fig:NMSE-versus-SNR}NMSE of radar/communication channel estimation
versus SNR.}
\end{figure}

Fig. \ref{fig:RMSE-versus-SNR} shows the root mean square error (RMSE)
of target/scatterer localization versus signal to noise ratio (SNR).
It is evident that the proposed SEA-Turbo-SBI based on joint estimation
outperforms the baselines in both target and scatterer localization.
The normalized mean square error (NMSE) of radar/communication channel
estimation versus SNR is shown in Fig. \ref{fig:NMSE-versus-SNR}.
It can be seen that the proposed SEA-Turbo-SBI also achieves the best
channel estimation performance. The performance gain between the proposed
SEA-Turbo-SBI based on joint estimation and separate estimation reflects
our proposed sparse prior model can fully exploit the partially common
sparsity of the location domain radar and communication channels.

\section{Conclusions}

We propose a location domain channel modeling method and a joint scattering
environment sensing and channel estimation scheme for a massive MIMO-OFDM
ISAC system. The SEA-Turbo-SBI algorithm is designed by combining
the turbo approach and the EM method, and the proposed sparse prior
model can exploit the partially common sparsity of the radar and communication
channels. Simulations verified that our proposed scheme can outperform
the baselines in both scattering environment sensing and channel estimation.

\bibliographystyle{IEEEtran}
\bibliography{Localization_CE,ISAC}

\end{document}